\documentstyle[aps,pre,multicol,epsf]{revtex}
\def\be{\begin{equation}}
\def\ee{\end{equation}}

\def\bea{\begin{eqnarray}}
\def\eea{\end{eqnarray}}

\begin{document}

\title{Formation of
Liesegang patterns: Simulations using a kinetic Ising model}
\vspace {1truecm}

\author{T. Antal${}^1$, M. Droz${}^1$, J. Magnin${}^{1,2}$, 
A. Pekalski${}^3$, and Z. R\'acz${}^{4}$}

\address{{}$^1$ {D\'epartement de Physique Th\'eorique, Universit\'e de 
Gen\`eve,
CH 1211 Gen\`eve 4, Switzerland.}}
\address{{}$^2$Center for Stochastic Processes in Science and 
Engineering and Department of Physics,\\ Virginia Tech, Blacksburg, VA 24061-0435, USA}
\address{{}$^3$Instytut Fizyki Teoretycznej, Uniwersytet Wroc{\l}awski, Wroc{\l}aw,
pl. M. Borna 9, 50-204 Wroc{\l}aw, Poland}
\address{{}$^4$Institute for Theoretical Physics,
E\"otv\"os University,
1117 Budapest, P\'azm\'any s\'et\'any 1/a, Hungary}

\maketitle
\begin{abstract}

A kinetic Ising model description of
Liesegang phenomena is studied using Monte Carlo
simulations. The model takes into account  
thermal fluctuations, contains noise in the chemical reactions,
and its control parameters are experimentally accessible.
We find that noisy, irregular
precipitation takes place in dimension $d=2$ while,
depending on the values of the control parameters, either
irregular patterns or precipitation bands satisfying 
the regular spacing law emerge in $d=3$.  

\end{abstract}

\date{\today}

\maketitle
\begin{multicols}{2}
\narrowtext

\section{Introduction}

Quasiperiodic precipitation patterns emerging 
in the wake of chemical reaction fronts are called
Liesegang patterns \cite{{liese},{Henisch}}. They have been studied for 
more than a century and a number of theoretical 
approaches have been developed to explain the experimental 
observations \cite{MP98}. 
Nevertheless, it was only recently that 
a model with input parameters fewer
than the number of static and dynamic parameters characterizing the 
patterns has appeared \cite{ADMRprl99}.

This last theory is based on the assumption
that the main ingredients of a macroscopic description 
should be a moving reaction front and the phase 
separation of the reaction product behind the front. Taking the 
properties of the reaction front from the theory of the 
fronts in the $A+B\rightarrow C$ process \cite{GR} and describing  
the phase separation process by the Cahn-Hilliard equation \cite{CahnHill}, 
one arrives at a model with a minimal number of parameters.

The above theory successfully explains that 
the positions, $x_n$ of the precipitation bands form 
a geometric series, $x_n\sim (1+p)^n$ (spacing law \cite{jabli}),
and gives the spacing coefficient $p$ in terms
of the initial concentrations of the reactants $A$ and $B$
in agreement with the Matalon-Packter law 
\cite{{Matalon},{Packter}}. Furthermore,
the parameters in the model can be determined from experiments and the
time-scale of the emergence of a band can be calculated \cite{ZR99}. 
Finally, the width law relating the position and the width of the bands 
can also be derived \cite{{ZR99},{JMthesis}} in agreement 
with observations \cite{widthlaw}.

The success and versatility notwithstanding, this theory needs 
further developments since, 
in its present form \cite{ADMRprl99}, it is a mean field theory 
without the fluctuations being accounted for. There are two ways 
to include the fluctuations. One is to add conserved thermal noise
to the Cahn-Hilliard equation as it is done in Model B of critical
dynamics \cite{HalpHoh}. In this case, there would be an additional 
problem of handling the noise in the reaction zone. It has been 
shown \cite{cornell}, however, that this noise is
irrelevant for $d>2$ so, in principle, it could be neglected.

In this work, another approach is taken for including the fluctuations. 
Namely, we shall study the kinetic Ising model version of the
process and  thus include the noise through the probabilistic 
description of the transitions between discrete states of the system. 

There are several reasons for our choice. First, the 
problem of difference between the handling of fluctuations in the
diffusive and reaction processes does not arise. Second,
the kinetic Ising version of the model has a meaning of 
a mesoscopic (or perhaps a microscopic) description.  
Since the mechanism of band formation may be at work
at a length-scale that makes possible the construction of 
submicron Liesegang structures, this kinetic Ising 
model approach may have a direct bearing on future experiments 
\cite{mikronexp}. Third, our choice was also influenced by 
having more expertise in simulations of kinetic Ising models. 

We shall start (Sec.II.) by a detailed discussion of the kinetic Ising model
designed to describe the band formation. 
The simulation results for this model are presented in Sec.III.
First, the $d=2$ case is treated where we do not find regular 
band formation. Then the $d=3$ simulations are discussed which
show the emergence of Liesegang patterns 
satisfying the usual spacing law. Final comments about the 
difficulties of predicting temperature-dependent effects in Liesegang
phenomena can be found in Sec.IV.
   
\section{Kinetic Ising model description}

\subsection{General aspects of the theory}

The aim of the theories of Liesegang phenomena is to explain how  
a high-concentration electrolyte $A$ diffuses into a gel soaked by a 
low-concentration electrolyte $B$ and how the spatial 
distribution of the precipitate $D$ is formed
in the wake of the diffusive reaction front. 
Accordingly, all the theories follow the scheme  
\be
A+B\rightarrow C\rightarrow D \, .
\label{A+B->C->D}
\ee
where $C$ is an intermediate 
reaction product which is generally not very well known. 
This uncertainty 
is then the basis for the existence of a number of competing theories 
\cite{{Ostw},{Wagn},{Chatt},{shino},{dee},{luthi},{postnuc},{reux}} with 
the differences arising from the interpretation of  
$C$ and from the level of details in the description of the
dynamics of $C$-s. 
A significant drawback of all these theories is that they contain 
a large number of parameters and many of them
are uncontrollable experimentally. Thus it is not entirely surprising that  
thorough comparisons between experiments and theories have not been 
carried out.  

\subsection{Cahn-Hilliard equation with source}

In our view, the uncertainty about the intermediate product and its 
dynamics can be used to advantage in building a general 
theory of band formation. One can interpret the 
concentration of $C$-s, $c$, as a kind of {\it order parameter} 
that takes a value $c_p$ 
in the ordered (precipitate) phase and another value $c_g$ in the disordered
(low-density) phase. The $C$-s are obtained from 
the $A+B\rightarrow C$ process so they are produced 
in the reaction zone. Furthermore, the dynamics of $C$-s obeys 
global conservation and it should be a phase-separation type 
dynamics since, in the expected final state, one has regions
of high- (precipitation) and low-density (interband) regions in 
equilibrium. This phase-separation dynamics can be described on 
a coarse-grained level by the Cahn-Hilliard equation \cite{CahnHill}
with the generation of $C$ appearing as an 
additional source term. The resulting equation for the space- and 
time-dependent 
order-parameter density, $c(x,t)$ is given by \cite{ADMRprl99}
\be
\partial_t c= -\lambda \Delta\left[\,\frac{\delta f(c)}{\delta c} +\sigma 
\Delta c\,\right] + S(x,t) \, .\label{move}
\ee
Here $\lambda$ is a kinetic coefficient, $f(c)$ is the Landau-Ginzburg 
free energy of the 
system which should have two equal minima at $c=c_p$ and $c=c_g$. The term 
$\sigma \Delta c$ with $\sigma>0$ provides stability against 
short-wavelength fluctuations, and finally $S(x,t)$ is 
the production rate of $C$-s 
in the reaction-diffusion process $A+B\to C$.
The properties of the source are known \cite{{GR},{koza},{Mag2000}}. 
It is localized, its   
center, $x_f$, moves diffusively ($x_f=\sqrt{2D_ft}$), and
it leaves behind a uniform concentration
$c_0$ of $C$-s. 

The parameters $\lambda$ and $\sigma$ in eq.(\ref{move}) 
can be used to set the 
time-scale and length-scale, respectively, and the source, $S$, is 
completely specified by the initial densities ($a_0,b_0$) 
and diffusion constants ($D=D_a\approx D_b$) of $A$ and $B$ 
\cite{{GR},{koza},{Mag2000}}.
Only the function $f(c)$ remains to be parametrized.    
One expects that the details of this function will not 
affect the overall properties of the pattern-formation process, the existence 
of two minima at $c_p$ and $c_g$ being the only important feature. Thus, 
assuming e.g. a coexistence curve that is symmetric about 
${\bar c}=(c_p+c_g)/2$,
one can parametrize this function as $f(c)=-\epsilon (c-{\bar c})^2 +
\gamma (c-{\bar c})^4$. The scale of the concentration can be set by 
the parameter $\gamma$ and  $\epsilon$ remains 
a free parameter in the theory. Consequently, one has a theory 
which has only a single adjustable parameter apart from the 
parameters setting the scale of the length, time, and the 
concentration field.

As discussed in the Introduction,  
all the observed general features 
of Liesegang phenomena can be derived from the 
above theory \cite{ADMRprl99},
including the time-scale of the emergence of a single band and the
length-scale for the width of the bands \cite{{ZR99},{JMthesis}}. 
Actually, it is somewhat surprising that the mean-field level description
in terms of eq.(\ref{move}) performs so well. The reason for this   
may lie in the experimental observation that the 
patterns are frozen (they do not evolve over time-scales 
extending up to 30 years 
\cite{Henisch}) meaning that the dynamics takes place at a very low
effective temperature, i.e.\ the noise is negligible.

The noise may indeed be negligible in the late stages of the formation of 
precipitation bands but the initial stages should be related 
to some instabilities and there the fluctuations should play a 
more prominent role. In particular, the mean-field description relies on 
a spinodal-decomposition instability (the moving front generates particles
and pushes the concentration past the spinodal) and the possibility
for the precipitation to take place through a nucleation-and-growth 
mechanism (where the fluctuations are important) is completely lost.

\subsection{Kinetic Ising model with Kawasaki + Glauber dynamics.}

In order to include fluctuations in the Liesegang process, 
a kinetic Ising model
will now be considered that is, we believe, a finite-temperature 
extension of the theory embodied in eq.(\ref{move}).  
The model introduced below can be viewed in two ways. Either it is a 
discretization scheme to equation (\ref{move}) and then the description 
is on a mesoscopic level with $c(x,t)$ being the discretized concentration  
of the order parameter, $C$.  Or, it can also be viewed as a simulation of the 
stochastic motion of $C$-s which are now particles at microscopic scales 
(in this case the coarse graining has been carried out in time).
In the following we shall use the latter "particle" language.

Let us begin the introduction of discretized description 
by identifying the particles $C$ 
with the up-spins of an Ising model on a hypercubic lattice. 
Then the down-spin sites are 
empty places and the formation of precipitation bands is modeled by a
combination of spin-flip and spin-exchange dynamics \cite{DRS}. Namely,
the initial state is prepared with all spins down 
(no $C$ particles present) and
the localized front flips the down spins (Glauber
dynamics~\cite{glauber}) thus producing the $C$-s. The flip rate 
$w_{\vec r}$ at site $\vec r$ is given by
\be
 w_{\vec r}=S(x,t)\frac{1}{2}(1-\sigma_{\vec r}) \, ,
 \label{fliprate}
\ee
where $\sigma_{\vec r}=\pm 1$ is the 
Ising spin at site $\vec r=(x,\vec r_{\perp})$ with $x$ being the 
coordinate in the 
direction of the motion of the reaction front while $\vec r_{\perp}$ 
representing the coordinates in the transverse direction (length 
is measured in units of the lattice spacing $a$). The 
factor  $(1-\sigma_{\vec r})$ ensures that the front flips only 
down spins (the particles are produced in the front
and the back reaction is negligible). Finally, $S(x,t)$ is a function 
describing the motion of the reaction front and the change of the 
reaction rate with time. The front is assumed to be homogeneous in
the transverse direction and its actual shape, $S(x,t)$, can be
taken from the solution and simulations of $A+B\rightarrow C$ process 
\cite{{GR},{larralde}}. Since the width of the reaction zone is small
and it changes with time very slowly, for all practical purposes, the 
function $S(x,t)$ can be approximated by a constant within a small interval 
$[x_f-\Delta,x_f+\Delta]$ around the center of the reaction 
zone $x_f=\sqrt{2D_ft}$
\be
S(x,t)=\frac{A}{\sqrt{t}}\theta{(x-x_f+\Delta)}\theta{(x_f+\Delta-x)} 
\label{S_i} .
\ee
where $\theta(x)$ is the step function and the amplitude 
\be
A=\frac{\sqrt{2D_f}}{4\Delta}c_0
\label{A}
\ee
is chosen such that the front leaves behind a constant 
$(c_0)$ concentration of particles \cite{{MP98},{GR}}. 

Once the particles are created, they diffuse and interact. 
This part of the dynamics can be described by a spin
exchange process (Kawasaki dynamics~\cite{kawa}). 
The rates of the exchanges
are assumed to satisfy detailed balance at temperature $T$ 
with ferromagnetic coupling ($J>0$) between the spins in order to 
describe the expected attraction among the $C$'s. Assuming the usual 
nearest-neighbor Ising Hamiltonian
\be
H=-J\sum_{<{\vec r},{\vec r'}>}\sigma_{\vec r}\sigma_{\vec r'} \, ,
\label{Isingham}
\ee
the rate of exchange between neighboring sites $\vec r$ and $\vec r^{\;\prime}$
can be chosen to be \cite{kawa}
\be
w_{{\vec r}\leftrightarrow {\vec r'}}=
\frac{1}{\tau_{e}}
\left [ 1+e^{\delta E/(k_BT)} \right ]^{-1}
\label{exchrate}
\ee
where $\tau_{e}$ sets the timescale, $T$ and $k_B$ are 
the temperature and the Boltzmann constant, respectively, 
and $\delta E$ is the change in the 
energy $\delta E = \delta H(\sigma_{\vec r}\leftrightarrow \sigma_{\vec r'})$ 
due to the exchange of spins at $\vec r$ and $\vec r^{\;\prime}$.

Without the spin-flip dynamics,
the system would relax to the equilibrium of the Ising model at 
temperature $T$ and at fixed magnetization. Thus choosing the 
temperature low enough (below $T_c$ of the Ising model) and making the 
spin-flip front produce the right magnetization density, the system 
will be in the unstable part of the phase diagram of the Ising model 
and phase separation will take place. If this model represents
the pattern forming process correctly then one expects the emergence 
of bands of up and down spins in the wake of the moving spin-flip front.


\section{Simulation results}
\label{simul}

\subsection{Parameters}
Monte-Carlo simulations of the kinetic Ising model described above have been 
performed in dimensions d=2 and 3. The dimension-specific properties
of the lattices used in the simulations will be given in the appropriate 
subsections below. Here we enumerate and discuss only those 
adjustable parameters which are used independently of dimension:

\begin{itemize}

\item The temperature $T$ is measured in units of $J/k_B$, where $J$ is the
nearest-neighbor coupling of the Ising model (\ref{Isingham}). It is 
clear from the considerations of the previous section that $T<T_c$ should be 
used.

\item The particle concentration, $c_0$, deposited by the front must be 
chosen so that, at the given $T$, it places the system in the metastable 
or unstable region of the phase diagram of the Ising model.  

\item Length is measured in units of the lattice spacing $a$. 
It should be noted that the value of $a$ depends on the interpretation 
of $C$ and $a$ can be
a microscopic length-scale as well as a mesoscopic one.

\item Time is measured in units of the "microscopic" 
timescale $\tau_e$ [see eq.(\ref{exchrate})].  
Again, this timescale may be coming 
from microscopic or mesoscopic processes depending  
on the interpretation of $C$.

\item The diffusion coefficient of the front ($D_f$) is, in principle, 
a well defined quantity \cite{GR}. The uncertainty of the connection 
between the microscopic and macroscopic length- and time-scales, however, 
makes it difficult to set a value for $D_f$ in terms of $a$ and $\tau_e$.
We shall thus treat $D_f$ as a parameter that can be freely varied.

\item The width of the reaction front ($2\Delta$) does not appear 
to influence the emerging patterns (this is the experience both from 
the mean-field theory \cite{ADMRprl99} and from small scale
simulations). Thus, in most of our simulations, we 
set $\Delta=1/2$ i.e.\ the front coincides with one of the lattice
planes perpendicular to the motion of the front.
\end{itemize}

The three important parameters are $T$, $c_0$ and $D_f$, and one should 
search for pattern formation in this three dimensional parameter space. 
This is not necessarily an easy task since 
it is known, both experimentally~\cite{Henisch} and theoretically~\cite{luthi},
that patterns are formed only in a restricted domain of the available 
free parameters. Since our search is finite, our 
statements (especially about the absence of patterns) are always 
pertinent only to the parameter domain investigated.

\subsection{\bf{Two-dimensional simulations}}

Simulations have been performed on stripes of length $L_x$ and width $L_y=50$ 
or $100$ with periodic boundary conditions in the transverse ($y$) direction. 
$L_x$ was chosen to be long enough so that no end effects would be observed 
with $L_x = 3000$ being a typical value. The initial position of
the front was always at $x=1$ and we used closed boundary conditions
(no particle crossing) at both $x=0$ and $x=L_x$. 

\begin{figure}[htb]
\centerline{
        \epsfxsize=9cm
        \epsfbox{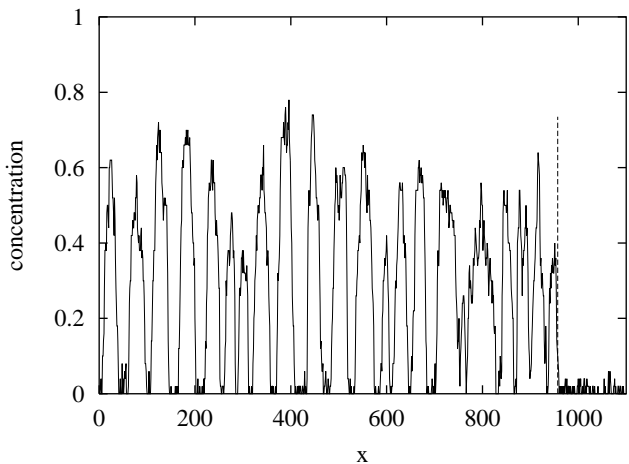}
           }
\vspace{0.5cm}
\caption{Concentration profile of $C$-s averaged over the 
transverse direction for a two-dimensional pattern 
obtained in a stripe of size $50 \times 3000$ with
$T=0.7, D_f=0.025$ and $c_0=0.3$. The distance along the slab ($x$)
is measured in units of the lattice spacing.
The dotted line shows the position of the front 
at the time ($t=1.8*10^7$) the
concentration was measured. The time is 
in units of the inverse of 
the rate of hopping for free $C$ particles ($\tau_e$). }
\label{Fig:d2random}
\end{figure}

The possible formation of patterns was investigated for the following 
ranges of the adjustable parameters: 
$0.02 \le T \le 1.5$, $0.01 \le D_f \le 1$
and $0.05 \le c_0 \le 0.5$. About 200 sets of values have been studied 
in the above domain and no regular patterns were observed. 
A typical result is displayed in 
Fig.\ref{Fig:d2random} for $T=0.7, D_f=0.025$ and $c_0=0.3$.

We have also made a few simulations outside of the above domain in order
to check the possible strong effect of the changes in a single 
parameter. These nonsystematic searches did not lead to pattern-forming
regimes either. 

We have extended the simulations to cases where the interactions are not 
restricted to nearest neighbors but extend up to seven lattice 
spacings. No patterns were found, although the randomness of the pattern 
slightly decreased when the range of the interaction was increased. 
This is expected as in
the limit of long-range interactions one should reach a continuum, noiseless
limit. Thus the Cahn-Hilliard description [eq.(\ref{move})] should 
apply and we should observe regular band formation.

\begin{figure}[htb]
\centerline{
        \epsfxsize=9cm
        \epsfbox{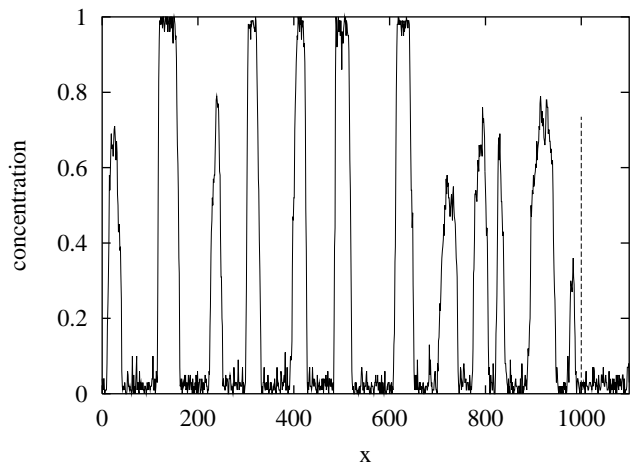}
           }
\vspace{0.5cm}
\caption{Concentration pattern in a system of size $50\times 2\times 3000$. 
The parameters $T$, $D_f$ and $c_0$ and the notations 
are the same as on Fig.\ref{Fig:d2random}.
The dotted line shows the position of the front.}
\label{Fig:d2+eps}
\end{figure}

The nonexistence of regular banding should not be considered 
as a contradiction with the experimental observations of $d=2$ Liesegang 
patterns. The experimental systems always have macroscopic width in the 
third dimension and this appears to stabilize the patterns. 
Indeed, Fig.\ref{Fig:d2+eps} shows a simulation with the same parameters 
as those in Fig.\ref{Fig:d2random},
except an extra layer in the third dimension is added. 
As one can see, the bands in the two-layer system are much better 
defined and they display some regularity.
Note in particular, that the concentration within the bands in 
Fig.\ref{Fig:d2+eps}
has reached the equilibrium value $(c\approx c_p\approx 1)$ 
while the maximum concentration regions in 
Fig.\ref{Fig:d2random} are roughly halfway in 
between the equilibrium values, $c_g\approx 0$ and $c_p\approx 1$.

\subsection{\bf Three-dimensional simulations}

The simulations were performed on slabs of length $L_x$ 
and of cross-section of size $L_y \times L_z$.
Periodic boundary conditions were used in the transverse 
($y,z$) directions while closed boundary conditions were employed 
at the two ends of the slabs. As in the two-dimensional case, $L_x$
was chosen so ($L_x \approx 3000$) 
as to avoid end effects from the transverse wall at $x=L_x$. 
Most of the simulations were done for $L_y=L_z\equiv L=10, 20$ and $40$ 
in order to observe finite-size effects. 
The initial condition was again an empty 
(all spins down) state with the front situated at $x=1$. 

\begin{figure}[htb]
\centerline{
        \epsfxsize=9cm
        \epsfbox{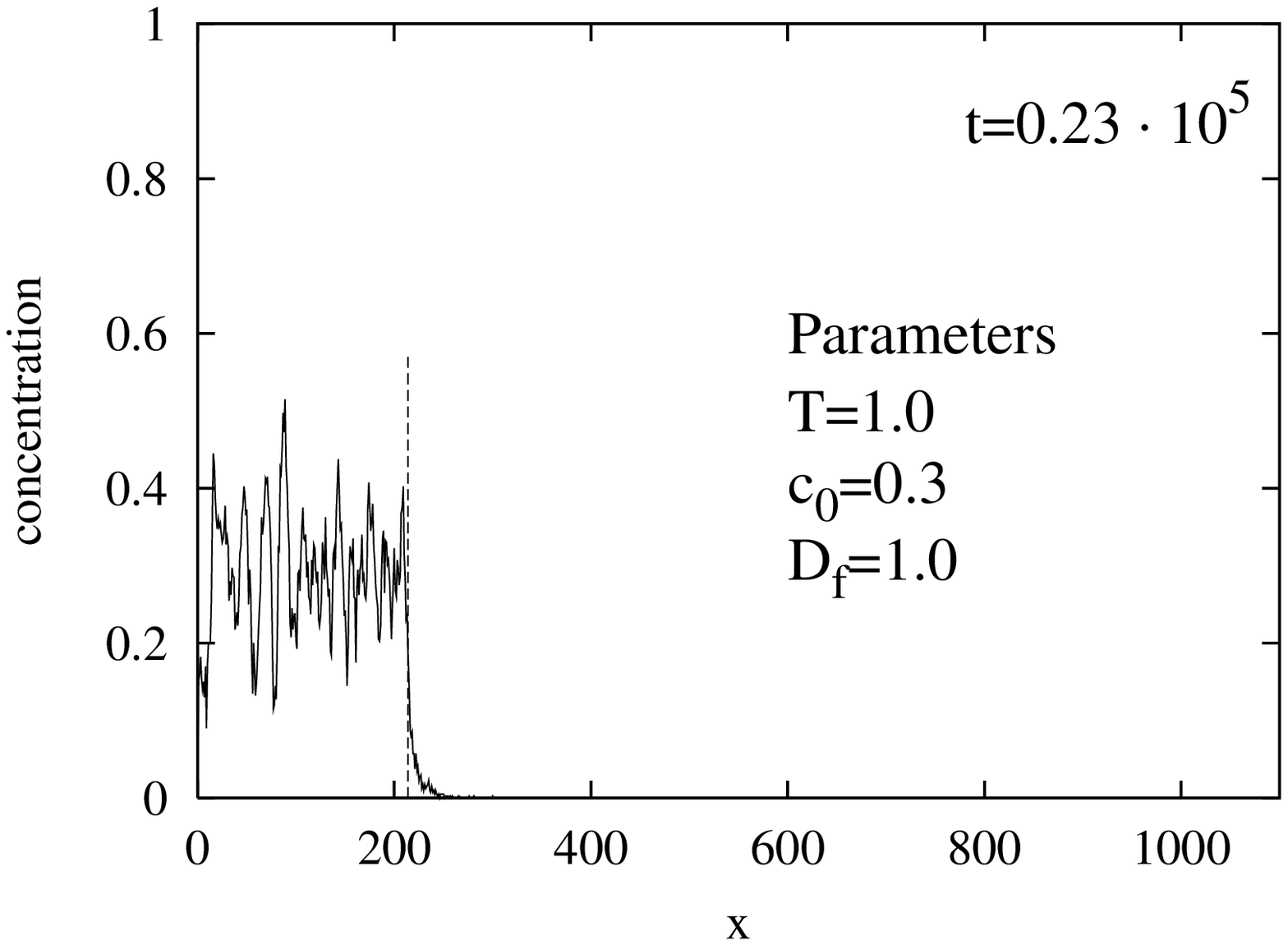}
           }
\vspace{-0.7cm}

\centerline{
        \epsfxsize=9cm
        \epsfbox{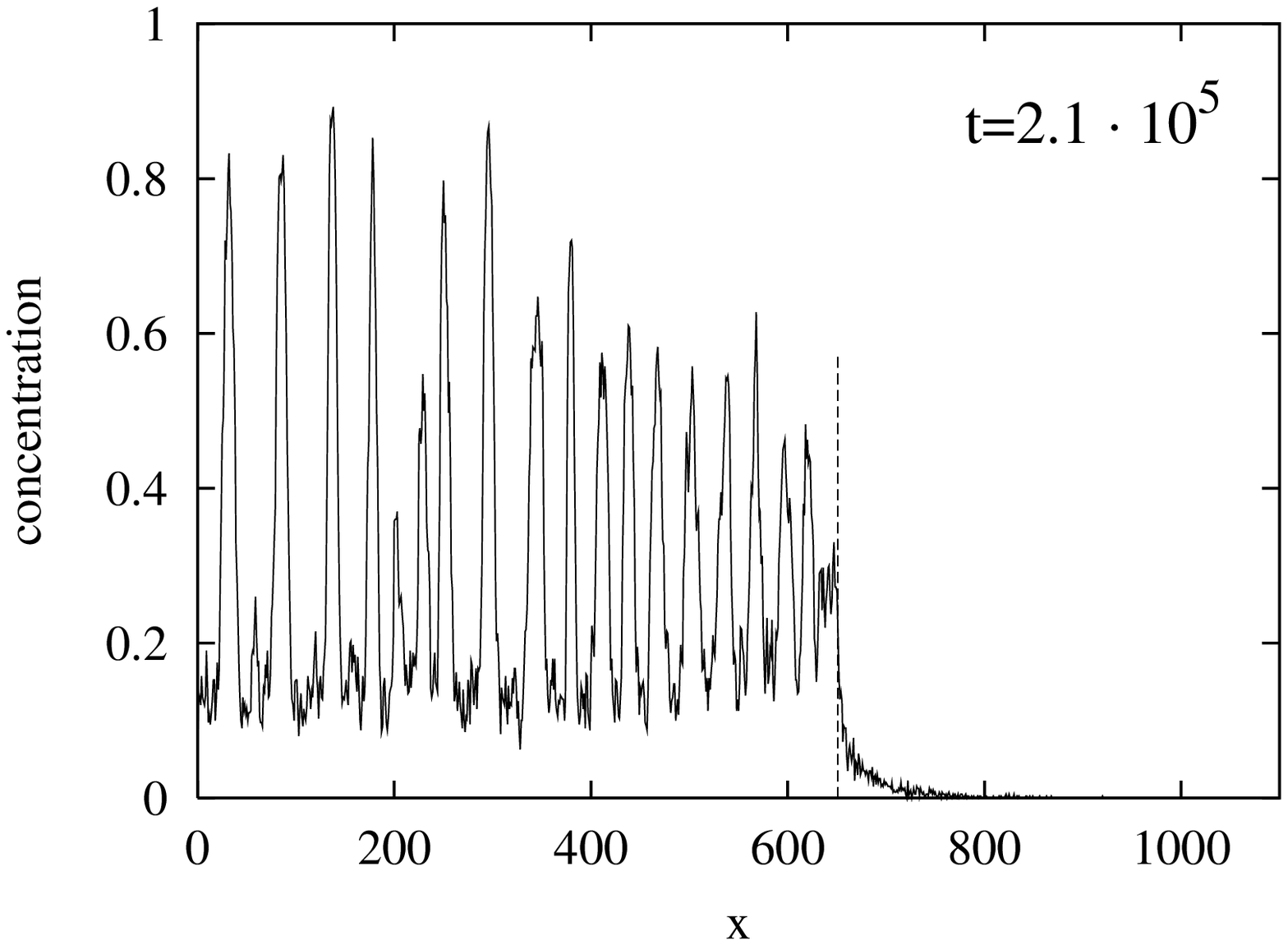}
           }
\vspace{-0.7cm}

\centerline{
        \epsfxsize=9cm
        \epsfbox{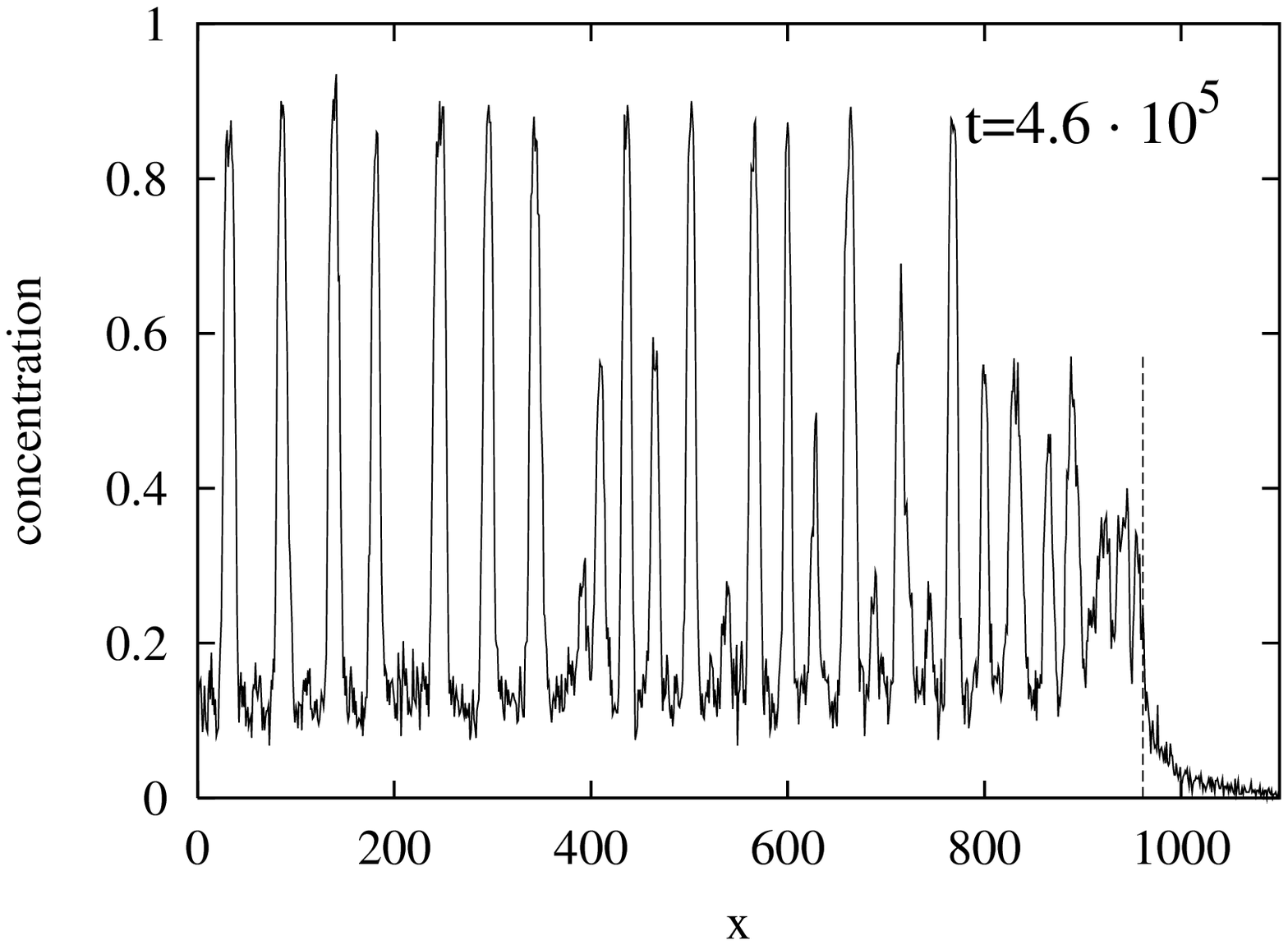}
           }
\vspace{0.5cm}\caption{Time evolution of the 
concentration of $C$ particles averaged over the transverse 
$(y,z)$ directions in slabs of size $20\times 20\times 3000$. 
The parameters ($T=1.0, D_f=1.0, c_0=0.4$) 
were chosen to be in the "coarsening pattern" regime.
The time is given in units of $\tau_e$ that is the inverse of 
the rate of hopping for free $C$ particles. 
The distance along the slab $(x)$ is measured in units of the 
lattice spacing. The dotted lines show the positions  
of the reaction front at time $t$.}
\label{Fig:d3random}
\end{figure}

Precipitation patterns were observed for $T\le 1$ if $c_0\ge 0.25$ was chosen 
well in the metastable or unstable region of the phase diagram 
of the d=3 Ising model. For large front diffusion ($D_f\ge 1$), 
these patterns were not regularly spaced and were not stable. 
Namely, coarsening was observed within reasonable observation time 
(time of formation of about 7-10 bands). A typical example is shown on
Fig.\ref{Fig:d3random}.

\begin{figure}[htb]
\centerline{
        \epsfxsize=9cm
        \epsfbox{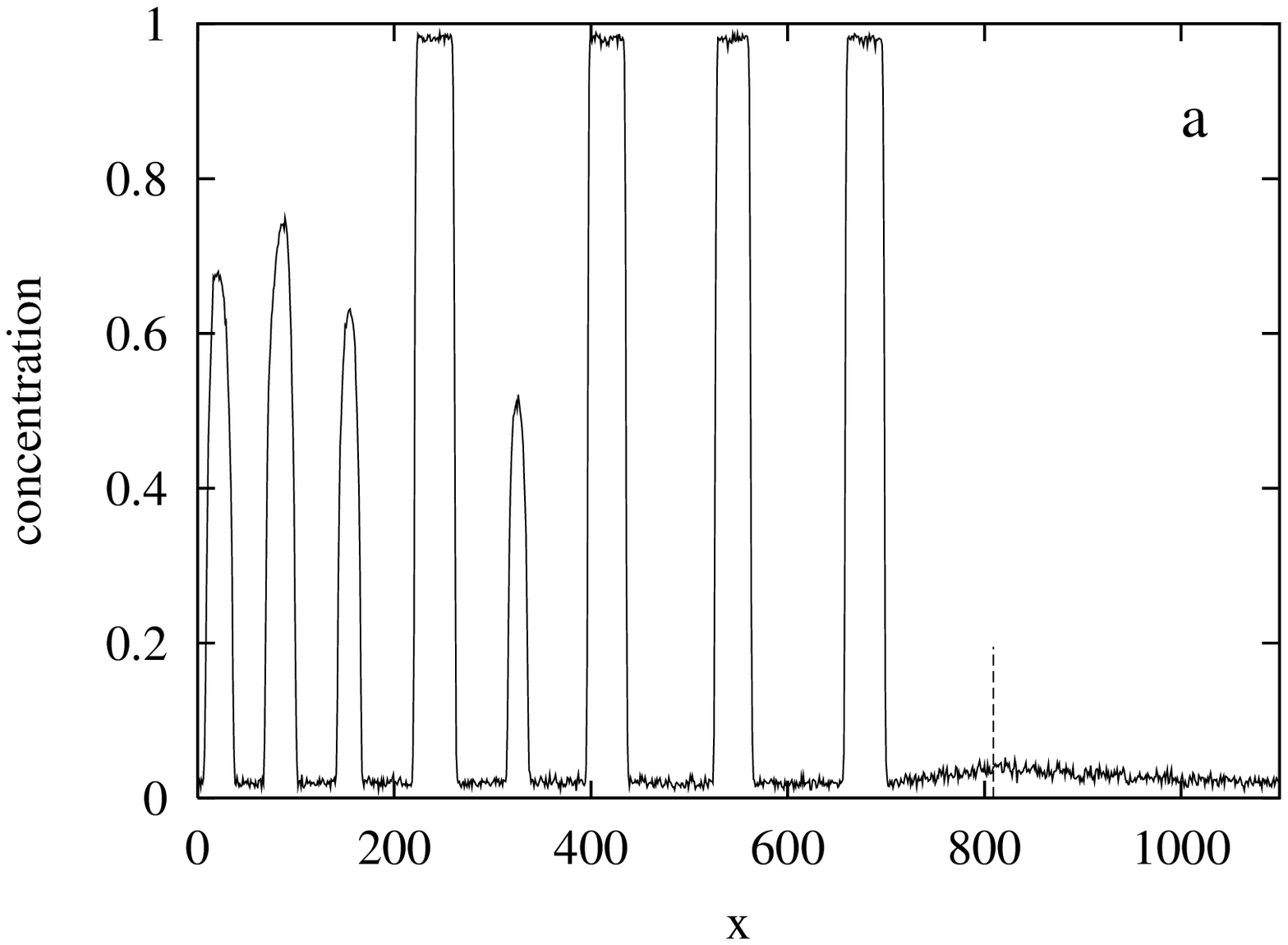}
           }
\vspace{-0.7cm}

\centerline{
        \epsfxsize=9cm
        \epsfbox{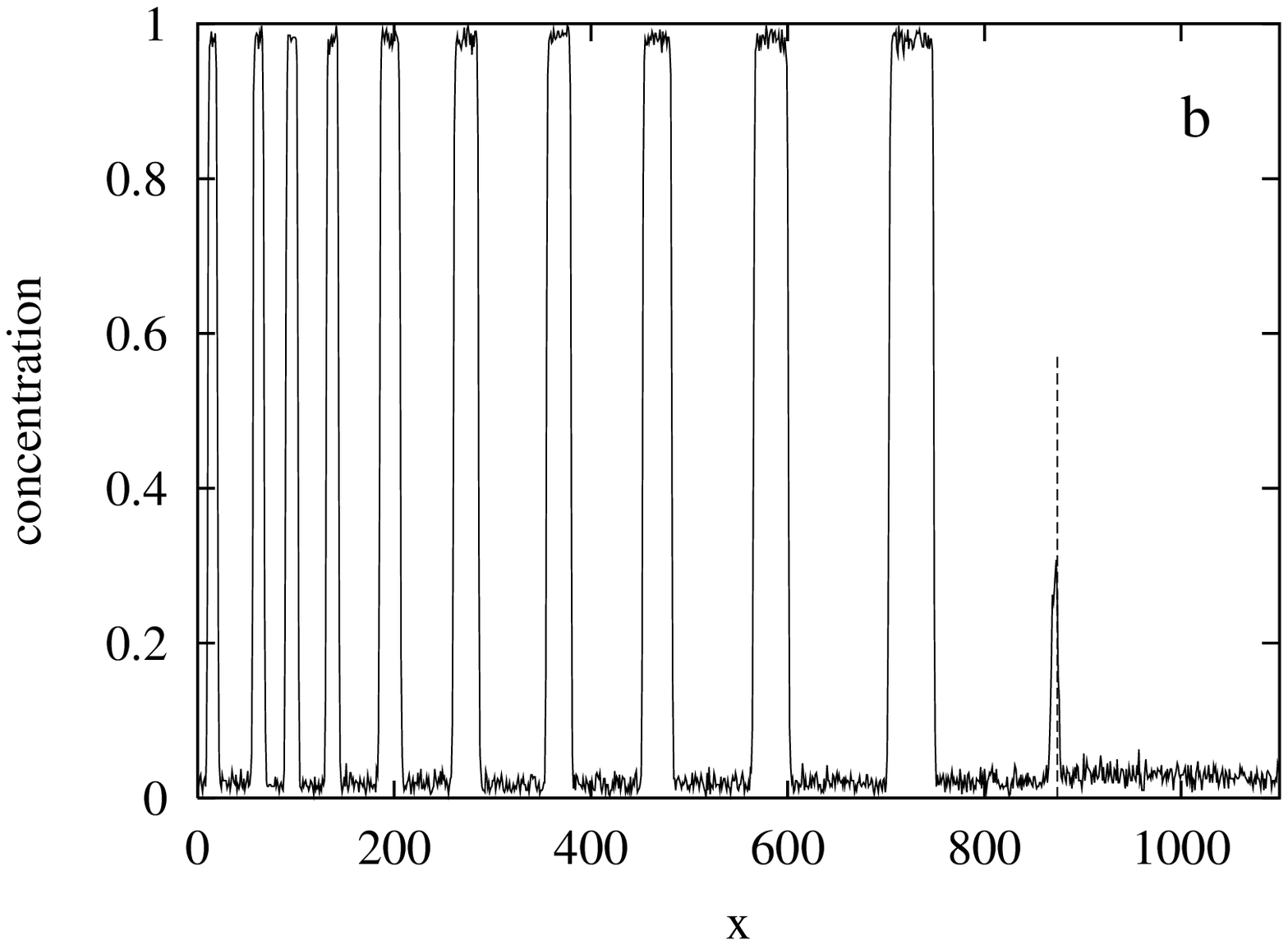}
           }
\vspace{-0.7cm}

\centerline{
        \epsfxsize=9cm
        \epsfbox{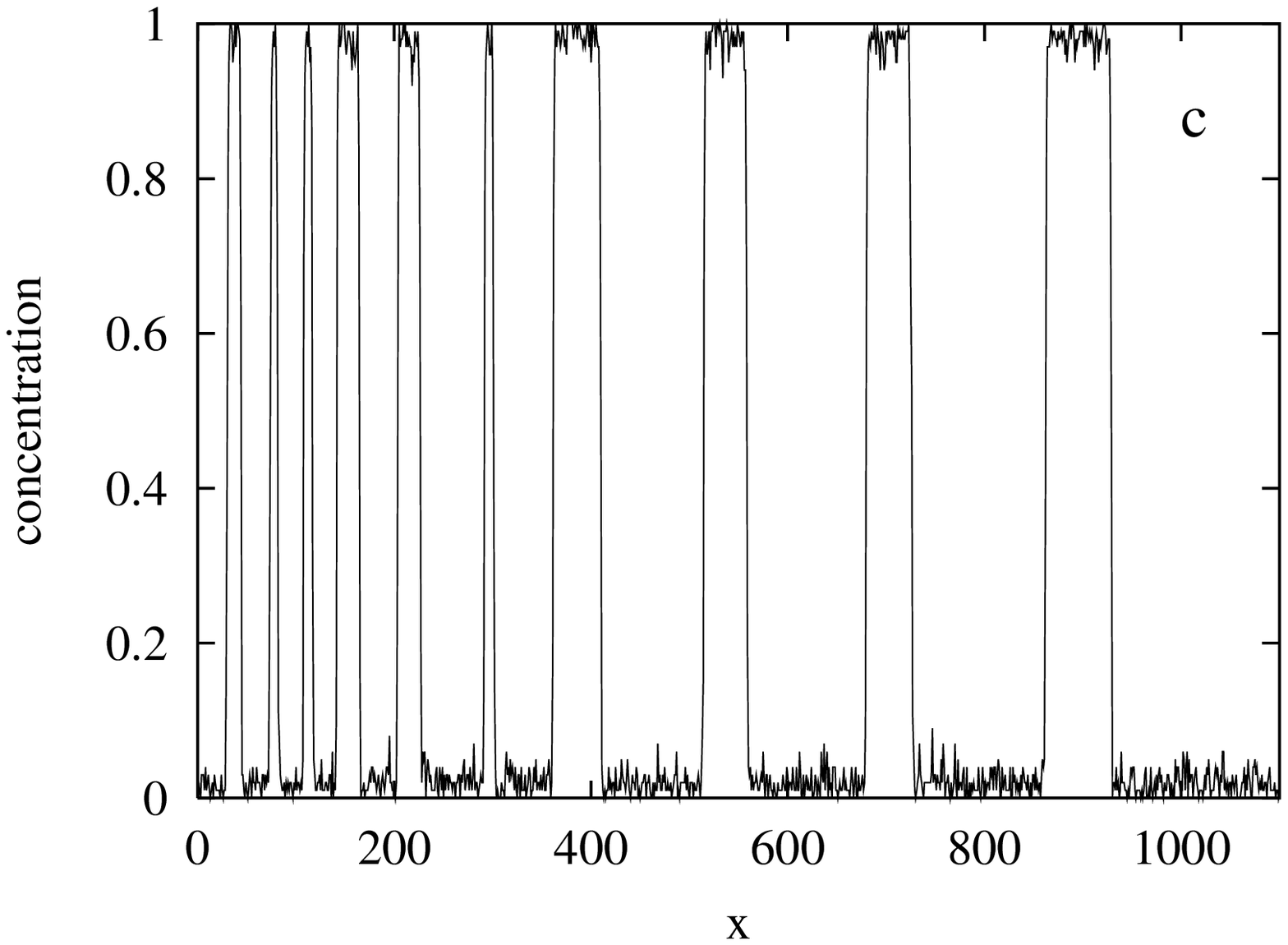}
           }
\vspace{0.5cm}
\caption{Concentration averaged over the transverse $(y,z)$ directions 
in slabs of size $L\times L\times 3000$. 
The values of the parameters 
($T=0.7, D_f=0.025$ and $c_0=0.3$) are the same for three transverse sizes 
$L=40$ (a), $L=20$ (b) and 
$L=10$ (c). The distance along the slab $(x)$ is measured in units of the 
lattice spacing. The dotted lines show the positions  
of the reaction front (the front is off the scale on 
Fig.\ref{Fig:d3liese}c).} 
\label{Fig:d3liese}
\end{figure}

The bands are more stable
at lower temperatures ($T\le 0.8$) and their spacing becomes more regular, 
as the front diffusion coefficient 
is decreased below $D_f\le 0.1$.
For  $D_f\le 0.05$, one observes the emergence of Liesegang type 
band patterns, typical examples being those shown 
on Fig.\ref{Fig:d3liese}a-c.
In this case, the system is of size $(L \times L \times 3000$) 
and the parameter values used are $T=0.7, D_f=0.025$ and $c_0=0.3$.
Results of runs for three cross sections ($L=10, 20$ and $40$) are 
displayed. Comparing these pictures, 
one cannot see any obvious finite-size trends. 

In order to investigate the spacing law one would need a large number
of bands. Unfortunately, in the regime where the best Liesegang type 
patterns are obtained one is restricted in the extent of explorations 
by the computing resources. 
Due to the  low value of $D_f$, the front is moving very 
slowly and consequently the computation time for obtaining 
e.g.\ 10 bands becomes very large. The CPU time necessary 
to produce the pattern on Fig.\ref{Fig:d3liese}a was about 1500 hours on a 
Sun Ultra-10 workstation. 

Fig.\ref{Fig:d3liese} shows roughly the limits of the 
possibilities of our simulations at present. To establish the spacing
law firmly, we would certainly need more bands. Using the last 
4-6 bands obtained 
from pictures similar to those on Fig.\ref{Fig:d3liese}, 
one can see that the positions 
of the bands ($x_n$) do approximate a geometric series $x_n\sim (1+p)^n$ 
well, and one can extract an approximate the spacing parameter, $p$.
In general, we find that the spacing coefficient 
does not show discernible finite-size trends. For example, 
the spacing parameter  
is $p\approx 0.17$ for all values of $L$ in Fig.\ref{Fig:d3liese}.

An interesting feature of the patterns emerging in our 
simulations is the presence of material ($C$) in between the 
bands (see Figures \ref{Fig:d3random}a-c and \ref{Fig:d3liese}a-c). 
Of course, it is not entirely clear whether part of 
this material should be considered as a low density precipitate 
(seen in many experiments \cite{Henisch}) or should it be 
just regarded as a "gas" phase of particles $C$. Visual inspection of 
the interband region reveals the presence of both small clusters  
and single particles. The small clusters live long
(especially at lower temperatures), their lifetime is 
comparable to the time of formation of several bands. Thus the
interpretation of part of the interband material as  low density 
precipitate may have some validity. In order to make progress in this
problem, one would need larger scale simulations as well as 
more understanding of the connection between the microscopic and
macroscopic time- and length-scales.

\section{Final remarks}

  Apart from the simplicity, an  
important feature of our model is that fluctuations are included. 
A frequent consequence of the presence of fluctuations is the 
disapperance of order in low dimensions and, indeed, our
simulations also indicate 
that Liesegang type patterns are absent in $d=2$ dimension while
they exist in $d=3$ dimensional samples.

Another advantage of including the noise is that the 
model has pattern-forming regimes which have 
characteristics of either the prenucleation theories \cite{{dee},{luthi}} or  
the postnucleation competitive growth theories \cite{postnuc}.
Indeed, one can see  (Fig.\ref{Fig:d3liese}b) that the bands are 
forming at the 
position of the reaction front for $D_f$ small, while the
bands are formed as a result of coarsening and competitive growth 
well behind the front in case of large $D_f$ (Fig.\ref{Fig:d3random}).
Note that the absence of Liesegang type patterns in the second regime 
is in agreement with the inability of producing 
such patterns in postnuclation competitive growth theories
\cite{postnuc}.

We feel that the most important feature of the model 
is that it makes clear that the important and experimentally controllable 
parameters are $T$, $c_0$ and $D_f$. Indeed, both $c_0$ and $D_f$ are 
known functions of the initial densities ($a_0,b_0$) 
and diffusion constants ($D=D_a\approx D_b$) of 
reagents $A$ and $B$ in the $A+B\rightarrow C$ reaction 
\cite{{GR},{koza},{Mag2000}}. Although the diffusion constants 
$D_a$ and $D_b$ are usually not controllable,
$a_0$ and $b_0$ can be set to given values and, consequently, 
$c_0$ and $D_f$ can be varied independently. 

Among the three parameters, a change in $T$ leads to unpredictable 
changes in the various diffusion coefficients and background processes
in a real Liesegang experiment 
(for example, the diffusion coefficients of the background ions
may also be important \cite{Unger} in determining the 
spacing coefficient). Thus it is advisable to keep the temperature  
constant, and the most promising and, indeed, most often 
used way of looking for trends in experiments \cite{{Matalon},{Packter}} 
is the change of 
the concentrations of the inner and outer electrolytes ($a_0$ and $b_0$). 

An obvious consequence of our simulations is that one should be able
to observe a crossover from prenucleation regime to 
a regime where postnucleation processes are dominant. Namely this 
could be achieved by tuning 
$a_0$ and $b_0$ in such a way that, at fixed $c_0$, the front diffusion
coefficient $D_f$ would be varied over as much a range as possible.

\section*{Acknowledgments}
This work has been supported by the 
Swiss National Science Foundation and 
by the Hungarian Academy of Sciences (Grant OTKA T 029792).

\end{multicols}

\end{document}